\documentclass[conference]{IEEEtran}
\IEEEoverridecommandlockouts
\usepackage{cite}
\usepackage{amsmath,amssymb,amsfonts}
\usepackage{algorithmic}
\usepackage{graphicx}
\usepackage{textcomp}
\usepackage{xcolor}
\usepackage{colortbl} 
\usepackage{booktabs} 
\usepackage{multirow} 
\usepackage{url}      
\usepackage{hyperref} 
\usepackage{float}
\usepackage{subcaption}
\usepackage{booktabs}
\usepackage{threeparttable}
\usepackage{tabularx}
\usepackage{caption}
\def\BibTeX{{\rm B\kern-.05em{\sc i\kern-.025em b}\kern-.08em
    T\kern-.1667em\lower.7ex\hbox{E}\kern-.125emX}}

\usepackage{xcolor}

\begin{document}

\title{Comparative Analysis of Lion and AdamW Optimizers for Cross-Encoder Reranking with MiniLM, GTE, and ModernBERT}

\author{\IEEEauthorblockN{Shahil Kumar}
\IEEEauthorblockA{\textit{Department of IT} \\ 
\textit{IIIT Allahabad}\\
Prayagraj, India \\
mml2023008@iiita.ac.in}
\and 
\IEEEauthorblockN{Manu Pande}
\IEEEauthorblockA{\textit{Department of IT} \\
\textit{IIIT Allahabad}\\
Prayagraj, India \\
mml2023005@iiita.ac.in}
\and 
\IEEEauthorblockN{Anay Yatin Damle}
\IEEEauthorblockA{\textit{Department of IT} \\
\textit{IIIT Allahabad}\\
Prayagraj, India \\
mml2023016@iiita.ac.in}
}


\maketitle

\begin{abstract}
Modern information retrieval systems often employ a two-stage pipeline consisting of an efficient initial retrieval stage followed by a more computationally intensive reranking stage. Cross-encoder models have demonstrated state-of-the-art effectiveness for the reranking task due to their ability to perform deep, contextualized analysis of query-document pairs. The choice of optimizer during the fine-tuning phase can significantly impact the final performance and training efficiency of these models. This paper investigates the impact of using the recently proposed Lion optimizer compared to the widely used AdamW optimizer for fine-tuning cross-encoder rerankers. We fine-tune three distinct transformer models, `microsoft/MiniLM-L12-H384-uncased`, `Alibaba-NLP/gte-multilingual-base`, and `answerdotai/ModernBERT-base`, on the MS MARCO passage ranking dataset using both optimizers. Notably, GTE and ModernBERT support longer context lengths (8192 tokens). The effectiveness of the resulting models is evaluated on the TREC 2019 Deep Learning Track passage ranking task and the MS MARCO development set (for MRR@10). Our experiments, facilitated by the Modal cloud computing platform for GPU resource management, show comparative results across three training epochs. ModernBERT trained with Lion achieved the highest NDCG@10 (0.7225) and MAP (0.5121) on TREC DL 2019, while MiniLM trained with Lion tied with ModernBERT with Lion on MRR@10 (0.5988) on MS MARCO dev. Additionally, Lion optimizer demonstrates superior GPU utilization efficiency across all models, achieving efficiency gains of 2.67\% to 10.33\% while maintaining competitive performance. We analyze the performance trends based on standard IR metrics, providing insights into the relative effectiveness of Lion versus AdamW for different model architectures and training configurations in the context of passage reranking.
The code for training and evaluation is available at \url{https://github.com/skfrost19/Cross-Encoder-Lion-vs-AdamW}, and the trained models are available on Huggingface Model Hub \url{https://huggingface.co/collections/skfrost19/rerenkers-681320776cfb45e44b18f5f1}.
\end{abstract}


\section{Introduction}
Modern Information Retrieval (IR) systems employ two-stage retrieve-and-rerank pipelines \cite{hu2019retrievereadrerankendtoend}, where efficient first-stage methods (BM25 \cite{robertson2009probabilistic}, dense retrieval \cite{karpukhin2020dense}) retrieve candidate documents, followed by sophisticated reranking models for precision optimization. Cross-encoder models based on transformer architectures \cite{devlin2019bert} have achieved state-of-the-art reranking performance \cite{nogueira2020passagererankingbert, Nogueira2020Document} through simultaneous query-document processing (`[CLS] query [SEP] document [SEP]`), enabling deep token-level interactions despite higher computational costs. Modern models like GTE \cite{li2023towards} and ModernBERT \cite{modernbert} support extended context lengths (8192 tokens) compared to efficiency-focused models like MiniLM \cite{wang2020minilm}.

Optimizer selection critically impacts transformer fine-tuning effectiveness. While AdamW \cite{loshchilov2019decoupled} remains widely adopted, the recently proposed Lion optimizer \cite{chen2023symbolic} (Evo\textbf{L}ved S\textbf{i}gn M\textbf{o}me\textbf{n}tum), derived through symbolic mathematics, demonstrates improved performance and memory efficiency in vision tasks.

This work investigates Lion's effectiveness for cross-encoder passage reranking compared to AdamW across three transformer models with varying characteristics. Our contributions are:
\begin{itemize}
    \item We fine-tune three distinct transformer-based cross-encoder models (`microsoft/MiniLM-L12-H384-uncased`, `Alibaba-NLP/gte-multilingual-base`, and `answerdotai/ModernBERT-base`) on the large-scale MS MARCO passage dataset \cite{DBLP:journals/corr/NguyenRSGTMD16} using both Lion and AdamW optimizers.
    \item We utilize different training configurations optimized for specific models, including a lower learning rate (2e-6) and a Cosine Annealing scheduler for ModernBERT, contrasting with a higher rate (2e-5) and no scheduler for MiniLM and GTE.
    \item We evaluate the performance of the fine-tuned models across three training epochs on the TREC 2019 Deep Learning (DL) Track \cite{craswell2020overview} benchmark and the MS MARCO development set \cite{DBLP:journals/corr/NguyenRSGTMD16}.
    \item We provide a comparative analysis of the optimizers' impact on reranking effectiveness using standard IR metrics (NDCG@10, MAP, MRR@10, Recall@10, R-Prec, P@10).
    \item We demonstrate that Lion optimizer achieves superior GPU utilization efficiency compared to AdamW across all tested models, with efficiency gains ranging from 2.67\% to 10.33\%, while maintaining competitive or superior performance metrics.
    \item We utilize the Modal cloud platform \cite{modal_labs} for efficient GPU resource management (3x NVIDIA L40S-48GB) and reproducible experimentation.
\end{itemize}
The rest of the paper is organized as follows: Section \ref{sec:related_work} discusses related work. Section \ref{sec:methodology} details the models, optimizers, and training approach. Section \ref{sec:experimental_setup} describes the experimental setup, datasets, and evaluation protocol. Section \ref{sec:results} presents and discusses the results. Finally, Section \ref{sec:conclusion} concludes the paper and suggests future work.

\section{Related Work}
\label{sec:related_work}

\subsection{Cross-Encoder Reranking}
BERT-based cross-encoders for document reranking were pioneered by Nogueira et al. \cite{nogueira2020passagererankingbert, Nogueira2020Document}, demonstrating superior performance over traditional IR models and bi-encoders through joint query-document processing. Subsequent advances explored diverse architectures \cite{Lin2021PretrainedTF}, training strategies \cite{gao2021complementing}, and efficiency improvements \cite{hofstatter2020improving}. Modern developments include efficiency-focused models like MiniLM \cite{wang2020minilm} (knowledge distillation), multilingual models like GTE \cite{li2023towards} supporting longer contexts, and architecturally enhanced models like ModernBERT \cite{modernbert} incorporating RoPE \cite{su2023roformerenhancedtransformerrotary} and Flash Attention \cite{dao2022flashattentionfastmemoryefficientexact} for 8192-token sequences. The `sentence-transformers` library \cite{reimers2019sentence} provides standardized frameworks for cross-encoder training and deployment.

\subsection{Optimizers for Deep Learning}
While SGD with momentum remains fundamental, adaptive methods like Adam \cite{kingma2017adam} combining momentum with adaptive learning rates dominate deep learning optimization. AdamW \cite{loshchilov2019decoupled} improves Adam by decoupling weight decay from adaptive mechanisms, enhancing generalization. The Lion optimizer \cite{chen2023symbolic} employs a simplified approach using momentum tracking and sign operations (`update = sign(momentum) * lr`), requiring less memory by avoiding second moment estimates. Lion's effectiveness, primarily demonstrated in vision and vision-language tasks, motivates its evaluation in NLP/IR domains.

\subsection{Evaluation Benchmarks}
MS MARCO \cite{DBLP:journals/corr/NguyenRSGTMD16} serves as the standard training benchmark for passage retrieval models, containing real Bing queries with human-judged relevance. TREC Deep Learning Tracks \cite{craswell2020overview, craswell2021overview} provide challenging test collections with sparse relevance judgments from pooled system results. We evaluate on TREC DL 2019 passage ranking using `trec\_eval` \cite{trec_eval_github} metrics and MS MARCO development set for MRR@10, providing comprehensive effectiveness assessment across different query scales.

\section{Methodology}
\label{sec:methodology}

\subsection{Cross-Encoder Architecture}
\begin{figure}[htbp]
\centering
\includegraphics[width=0.9\linewidth]{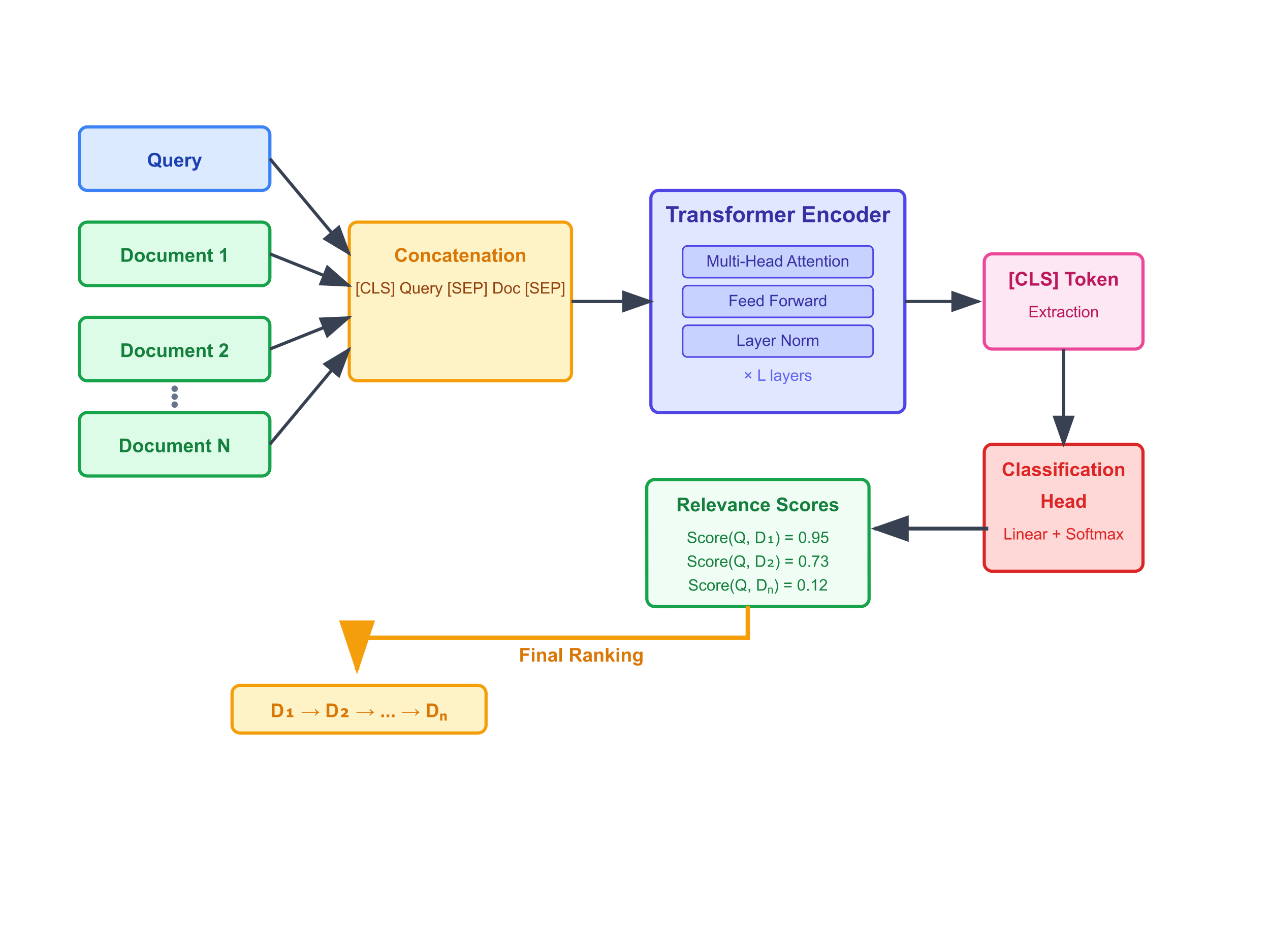}
\caption{General Cross-Encoder Architecture for Reranking.}
\label{fig:cross_encoder_arch}
\end{figure}

A cross-encoder model takes a query $q$ and a document $d$ as input, typically concatenating them with special tokens: `[CLS] q [SEP] d [SEP]`. This combined sequence is fed into a transformer model (e.g., MiniLM, GTE, ModernBERT). The output representation corresponding to the `[CLS]` token is then passed through a linear layer followed by a sigmoid activation to predict a relevance score $s(q, d) \in [0, 1]$. During inference for reranking, this score is computed for all candidate documents retrieved in the first stage, and the documents are re-sorted based on these scores. An overview of the architecture is shown in Fig. \ref{fig:cross_encoder_arch}.

\subsection{Base Models}
We experiment with three transformer base models from Hugging Face \cite{wolf2020transformers}:
\begin{itemize}
    \item \textbf{`microsoft/MiniLM-L12-H384-uncased` \cite{wang2020minilm}:} A distilled version of BERT, designed to be smaller (12 layers, 384 hidden size) and faster while retaining significant performance. It has a standard context length, typically 512 tokens.
    \item \textbf{`Alibaba-NLP/gte-multilingual-base` \cite{li2023towards}:} Part of the General Text Embeddings (GTE) family, trained on a large, diverse multilingual corpus. While often used as a bi-encoder, its underlying transformer architecture can be effectively fine-tuned as a cross-encoder. This 'base' version supports a context length of 8192 tokens.
    \item \textbf{`answerdotai/ModernBERT-base` \cite{modernbert}:} A state-of-the-art encoder-only transformer incorporating architectural enhancements like Rotary Positional Embeddings (RoPE) \cite{su2023roformerenhancedtransformerrotary} for effective long context handling (up to 8192 tokens), Flash Attention \cite{dao2022flashattentionfastmemoryefficientexact} for speed and memory efficiency, and GeGLU activation functions \cite{shazeer2020gluvariantsimprovetransformer}. It was trained on 2 trillion tokens and demonstrates strong performance on various tasks.
\end{itemize}

\subsection{Training}
We fine-tune the cross-encoders using the MS MARCO passage ranking triplets dataset \cite{DBLP:journals/corr/NguyenRSGTMD16}, as processed by `sentence-transformers` \cite{reimers2019sentence}. The original dataset contains tuples of (query, positive passage, negative passage). We convert this into pairs `(query, positive\_passage)` with label 1 and `(query, negative\_passage)` with label 0. We train on approximately 2 million such pairs.
The training objective is to minimize the Binary Cross-Entropy (BCE) loss between the predicted relevance score $\hat{y}$ (output of the sigmoid function) and the true label $y$ (0 or 1). The BCE loss is defined as:
\[ \mathcal{L}_{BCE} = - [y \log(\hat{y}) + (1 - y) \log(1 - \hat{y})] \]

\subsection{Optimizers}
We compare two optimizers using specified parameters:
\begin{itemize}
    \item \textbf{AdamW \cite{loshchilov2019decoupled}:} Implemented via \texttt{torch.optim.AdamW}. We use a weight decay of 0.01. The learning rate varies by model (see Section \ref{sec:experimental_setup}).
    \item \textbf{Lion \cite{chen2023symbolic}:} Implemented via a PyTorch implementation based on the original paper. We use the default betas $\beta_1 = 0.9$, $\beta_2 = 0.99$, and a weight decay of 0.01. The learning rate varies by model. The update rule is as follows:
    \begin{align*}
    c_t &= \beta_1 m_{t-1} + (1 - \beta_1) g_t \\
    \theta_t &= \theta_{t-1} - \eta \left( \text{sign}(c_t) + \lambda \theta_{t-1} \right) \\
    m_t &= \beta_2 m_{t-1} + (1 - \beta_2) g_t
    \end{align*}
    where $g_t$ is the gradient, $m_t$ is the momentum, $\eta$ is the learning rate, $\beta_1, \beta_2$ are momentum coefficients, and $\lambda$ is the weight decay.
\end{itemize}

\section{Experimental Setup}
\label{sec:experimental_setup}

\subsection{Datasets}
\begin{itemize}
    \item \textbf{Training:} We use the MS MARCO passage ranking triplets dataset \cite{DBLP:journals/corr/NguyenRSGTMD16}, processed into approximately 2 million query-passage pairs with binary labels using the `sentence-transformers` methodology.
    \item \textbf{Evaluation (Main):} We evaluate on the TREC 2019 Deep Learning Track passage ranking task \cite{craswell2020overview}, which contains 43 queries with graded relevance judgments (qrels).
    \item \textbf{Evaluation (MRR@10):} We evaluate Mean Reciprocal Rank at cutoff 10 (MRR@10) on the MS MARCO passage dataset's development split (available from Hugging Face Datasets \cite{DBLP:journals/corr/NguyenRSGTMD16}), containing a larger set of queries with binary relevance.
\end{itemize}

\subsection{Implementation Details}
\begin{itemize}
    \item \textbf{Framework:} We use `sentence-transformers` \cite{reimers2019sentence} (`CrossEncoder`) built on PyTorch \cite{paszke2019pytorchimperativestylehighperformance} and Hugging Face Transformers \cite{wolf2020transformers}.
    \item \textbf{Hyperparameters:} Key training parameters are:
        \begin{itemize}
            \item Batch Size: 64.
            \item Learning Rate:
                \begin{itemize}
                    \item 2e-5 for MiniLM and GTE (both optimizers).
                    \item 2e-6 for ModernBERT with Lion optimizer.
                    \item 2e-5 for ModernBERT with adamW optmizer.
                \end{itemize}
            \item Scheduler:
                \begin{itemize}
                    \item None for MiniLM and GTE.
                    \item CosineAnnealingLR Scheduler for ModernBERT, with `T\_max` set to the total number of training steps.
                \end{itemize}
            \item Optimizer Params:
                \begin{itemize}
                    \item AdamW: weight\_decay=0.01.
                    \item Lion: betas=(0.9, 0.99), weight\_decay=0.01.
                \end{itemize}
            \item Warmup Ratio: 0.1 (for LR scheduler if used)
            \item Epochs: 3
            \item Precision: BF16 enabled.
            \item Seed: 12 (for reproducibility).
            \item Dataloader Workers: 4.
        \end{itemize}
    \item \textbf{Infrastructure:} 
    \begin{itemize}
        \item Experiments were conducted using the Modal platform \cite{modal_labs}, leveraging 3x NVIDIA L40S-48GB GPUs. Modal managed the environment (CUDA, Python 3.11, libraries) and distributed training.

        \item Experiment tracking, visualization of training metrics (such as loss and learning rate), and hyperparameter logging were managed using the Weights \& Biases (W\&B) platform \cite{wandb2020}.
    \end{itemize}

    \item \textbf{First-Stage Retrieval:} For TREC DL evaluation, we retrieve the top 1000 candidate passages per query using a standard BM25 baseline via Pyserini \cite{lin2021pyserini} on the `msmarco-v1-passage` index. The cross-encoders rerank this candidate set.
\end{itemize}

\subsection{Evaluation Protocol}
\begin{itemize}
    \item \textbf{Reranking:} Each fine-tuned cross-encoder model scores the top-1000 BM25 candidates for each TREC DL 2019 query. Passages are reranked based on these scores.
    \item \textbf{Metrics:} We use `trec\_eval` \cite{trec_eval_github} for TREC DL 2019 evaluation. We report:
        \begin{itemize}
            \item NDCG@10 (Normalized Discounted Cumulative Gain @10)
            \item MAP (Mean Average Precision)
            \item Recall@10 (Recall @10)
            \item R-Prec (R-Precision)
            \item P@10 (Precision @10)
        \end{itemize}
    For MRR@10, we evaluate on the MS MARCO development set.
    \item \textbf{Configurations Tested:} We evaluate each combination of base model and optimizer after each of the 3 training epochs:
        \begin{itemize}
            \item MiniLM + AdamW (Epochs 1, 2, 3)
            \item MiniLM + Lion (Epochs 1, 2, 3)
            \item GTE-multilingual-base + AdamW (Epochs 1, 2, 3)
            \item GTE-multilingual-base + Lion (Epochs 1, 2, 3)
            \item ModernBERT-base + AdamW (Epochs 1, 2, 3)
            \item ModernBERT-base + Lion (Epochs 1, 2, 3)
        \end{itemize}
\end{itemize}

\section{Results and Discussion}
\label{sec:results}
Table \ref{tab:main_results} presents the evaluation results for all configurations on the TREC 2019 Deep Learning Track (NDCG@10, MAP, Recall@10, R-Prec, P@10) and MS MARCO development set (MRR@10) after 1, 2, and 3 epochs of fine-tuning. Best results for each metric are highlighted.

\begin{table*}[htbp]
\centering
\caption{Evaluation Results on TREC-DL 2019 and MS-MARCO Dev Passage Ranking}
\label{tab:main_results}
\begin{tabular}{l l c c c c c c c}
\toprule
\textbf{Base Model} & \textbf{Optimizer} & \textbf{Epoch} & \textbf{NDCG@10} & \textbf{MAP} & \textbf{MRR@10} & \textbf{Recall@10} & \textbf{R-Prec} & \textbf{P@10} \\
\midrule
\multirow{3}{*}{MiniLM-L12-H384} & \multirow{3}{*}{AdamW}
    & 1 & 0.7008 & 0.4814 & 0.5828 & 0.1712 & 0.4899 & 0.8047 \\
    & & 2 & 0.7094 & 0.4891 & 0.5818 & 0.1715 & 0.5017 & 0.8093 \\
    & & 3 & 0.7127 & 0.4908 & 0.5826 & 0.1706 & 0.4962 & 0.8023 \\
\midrule
\multirow{3}{*}{MiniLM-L12-H384} & \multirow{3}{*}{Lion}
    & 1 & 0.7031 & 0.4858 & 0.5890 & 0.1698 & 0.4904 & 0.8070 \\
    & & 2 & 0.6916 & 0.4755 & 0.5942 & 0.1724 & 0.5041 & 0.8116 \\
    & & 3 & 0.6808 & 0.4706 & \cellcolor{yellow!50}\textbf{0.5988} & 0.1701 & 0.4923 & 0.8023 \\
\midrule
\multirow{3}{*}{GTE-multilingual-base} & \multirow{3}{*}{AdamW}
    & 1 & 0.7224 & 0.5005 & 0.5940 & 0.1733 & 0.4957 & 0.8140 \\
    & & 2 & 0.7203 & 0.4999 & 0.5942 & \cellcolor{yellow!50}\textbf{0.1733} & 0.5067 & 0.8163 \\
    & & 3 & 0.6902 & 0.4899 & 0.5972 & 0.1730 & 0.5069 & 0.8140 \\
\midrule
\multirow{3}{*}{GTE-multilingual-base} & \multirow{3}{*}{Lion}
    & 1 & 0.6785 & 0.4754 & 0.5854 & 0.1684 & 0.4849 & 0.7953 \\
    & & 2 & 0.6909 & 0.4921 & 0.5957 & 0.1721 & 0.5053 & 0.8140 \\
    & & 3 & 0.6904 & 0.4912 & 0.5931 & 0.1719 & 0.5041 & 0.8093 \\
\midrule
\multirow{3}{*}{Modern-BERT-base} & \multirow{3}{*}{AdamW}
    & 1 & 0.7105 & 0.5066 & 0.5865 & 0.1678 & 0.5161 & 0.8163 \\
    & & 2 & 0.6839 & 0.4893 & 0.5885 & 0.1634 & 0.4946 & 0.7814 \\
    & & 3 & 0.6959 & 0.4971 & 0.5916 & 0.1623 & 0.5116 & 0.7860 \\ 
\midrule
\multirow{3}{*}{Modern-BERT-base} & \multirow{3}{*}{Lion}
    & 1 & 0.7142 & \cellcolor{yellow!50}\textbf{0.5121} & 0.5834 & 0.1689 & 0.5148 & 0.8163 \\
    & & 2 & \cellcolor{yellow!50}\textbf{0.7225} & 0.5115 & 0.5907 & 0.1732 & \cellcolor{yellow!50}\textbf{0.5183} & 0.8209 \\
    & & 3 & 0.7051 & 0.5020 & \cellcolor{yellow!50}\textbf{0.5988}* & 0.1722 & 0.5102 & \cellcolor{yellow!50}\textbf{0.8256} \\
\bottomrule
\end{tabular}
\vspace{1em}\\
\footnotesize{* MRR@10 is calculated on the MS MARCO v1.1 passage dataset development split. All other metrics are calculated on TREC DL 2019. Best result for each metric highlighted. (*Note: MiniLM+Lion also achieved 0.5988 MRR@10 at epoch 3.}
\end{table*}

\begin{table*}[htbp]
\centering
\caption{GPU Usage Statistics Comparison: Lion vs AdamW Optimizers}
\label{tab:gpu_usage}
\begin{tabular}{l l c c c c c}
\toprule
\textbf{Model} & \textbf{Optimizer} & \textbf{Mean Usage (\%)} & \textbf{Peak Usage (\%)} & \textbf{Std Dev (\%)} & \textbf{Data Points} & \textbf{Efficiency Gain (\%)} \\
\midrule
\multirow{2}{*}{MiniLM-L12-H384} 
    & AdamW & 33.09 & 35.25 & 2.72 & 410 & \multirow{2}{*}{\cellcolor{green!20}\textbf{+2.67\%}} \\
    & \cellcolor{green!20}Lion & \cellcolor{green!20}\textbf{32.21} & \cellcolor{green!20}\textbf{34.64} & 2.78 & 412 & \\
\midrule
\multirow{2}{*}{GTE-multilingual-base} 
    & AdamW & 73.04 & 78.08 & 7.05 & 698 & \multirow{2}{*}{\cellcolor{green!20}\textbf{+10.33\%}} \\
    & \cellcolor{green!20}Lion & \cellcolor{green!20}\textbf{65.50} & \cellcolor{green!20}\textbf{69.69} & 6.27 & 699 & \\
\midrule
\multirow{2}{*}{ModernBERT-base} 
    & AdamW & 77.04 & 81.40 & 6.01 & 578 & \multirow{2}{*}{\cellcolor{green!20}\textbf{+3.49\%}} \\
    & \cellcolor{green!20}Lion & \cellcolor{green!20}\textbf{74.35} & \cellcolor{green!20}\textbf{79.45} & 8.97 & 579 & \\
\bottomrule
\end{tabular}%
\vspace{1em}\\
\footnotesize{GPU utilization measured across 3 NVIDIA L40S-48GB GPUs during training. Efficiency gain calculated as: $\frac{\text{AdamW Mean} - \text{Lion Mean}}{\text{AdamW Mean}} \times 100\%$. Lower usage indicates better efficiency.}
\end{table*}

\subsection{GPU Efficiency Analysis}
Beyond performance metrics, we analyze the GPU resource efficiency of both optimizers during training. Table \ref{tab:gpu_usage} presents detailed GPU utilization statistics collected across 3 NVIDIA L40S-48GB GPUs during the training process.

The results demonstrate that Lion optimizer consistently achieves better GPU utilization efficiency compared to AdamW across all tested models:

\begin{itemize}
    \item \textbf{MiniLM-L12-H384:} Lion shows a modest efficiency improvement of 2.67\%, reducing mean GPU usage from 33.09\% to 32.21\% and peak usage from 35.25\% to 34.64\%.
    \item \textbf{GTE-multilingual-base:} The most significant efficiency gain of 10.33\% is observed with Lion, reducing mean GPU usage from 73.04\% to 65.50\% and peak usage from 78.08\% to 69.69\%.
    \item \textbf{ModernBERT-base:} Lion achieves a 3.49\% efficiency improvement, reducing mean GPU usage from 77.04\% to 74.35\% and peak usage from 81.40\% to 79.45\%.
\end{itemize}

These efficiency gains are particularly notable given that Lion maintains competitive or superior performance metrics (as shown in Table \ref{tab:main_results}). The lower GPU utilization suggests that Lion's simpler update rule, which only requires tracking momentum and does not store second moment estimates like AdamW, translates to practical computational savings during training. This aligns with the theoretical memory efficiency advantages of Lion optimizer reported in the original paper \cite{chen2023symbolic}.

The largest efficiency gain observed with GTE-multilingual-base (10.33\%) is especially significant considering this model's larger size and 8K context length capability. This suggests that Lion's benefits may be more pronounced for larger, more computationally intensive models.

\subsection{Optimizer Performance Comparison}
We observe different interactions between the base models and the optimizers:

\begin{itemize}
    \item **MiniLM:** AdamW generally achieves higher peak performance on TREC DL 2019 metrics (NDCG@10: 0.7127 vs 0.7031; MAP: 0.4908 vs 0.4858), peaking at Epoch 3. Lion achieves the overall best MRR@10 on MS MARCO dev (0.5988 at Epoch 3), but its TREC performance peaks earlier (Epoch 1) and then declines. This suggests AdamW might be more stable over longer training for MiniLM on TREC, while Lion might converge faster or optimize differently for the MS MARCO distribution.
    \item **GTE:** AdamW clearly outperforms Lion for the GTE model across most metrics and epochs. GTE+AdamW achieves its best TREC performance (NDCG@10=0.7224, MAP=0.5005) early at Epoch 1 or 2, with performance degrading by Epoch 3. GTE+Lion shows weaker performance overall, peaking later (Epoch 2/3) but never reaching the levels of GTE+AdamW.
    \item **ModernBERT:** Lion significantly outperforms AdamW when training ModernBERT with the specific low learning rate (2e-6) and Cosine Annealing scheduler. ModernBERT+Lion achieves the overall best NDCG@10 (0.7225), MAP (0.5121), R-Prec (0.5183), and P@10 (0.8256) in our experiments. Its performance peaks around Epoch 2 for most TREC metrics. ModernBERT+AdamW, using the same LR and scheduler, performed notably worse, suggesting Lion interacts more favorably with this setup for this model.
\end{itemize}
These results indicate that the choice of optimizer interacts significantly with the base model architecture and the specific training hyperparameters (learning rate, scheduler). Lion showed particular strength with ModernBERT under its tailored training regime.

\subsection{Model Performance Comparison}
Comparing the best results achieved by each model (irrespective of optimizer epoch):

\begin{itemize}
    \item **ModernBERT (+Lion, E2/E1/E3):** Achieved the highest NDCG@10 (0.7225), MAP (0.5121), R-Prec (0.5183), and P@10 (0.8256) on TREC DL 2019, and tied for the best MRR@10 (0.5988) on MS MARCO dev. This suggests its advanced architecture and long context, combined with the Lion optimizer and specific training strategy, yield superior reranking quality.
    \item **GTE (+AdamW, E1/E2):** Achieved the next best performance, with NDCG@10 (0.7224) nearly matching ModernBERT+Lion, and achieving the best Recall@10 (0.1733). Its peak performance was reached earlier than ModernBERT's in terms of epochs.
    \item **MiniLM (+AdamW E3 / +Lion E3):** While performing competitively, MiniLM generally lagged behind GTE and ModernBERT on TREC metrics. However, MiniLM+Lion achieved the joint-highest MRR@10 (0.5988) on the MS MARCO dev set, indicating strong performance on that specific benchmark despite lower TREC scores compared to the larger models.
\end{itemize}
The results suggest that the larger models with longer context capabilities (GTE, ModernBERT) generally outperform MiniLM on the TREC DL 2019 task. ModernBERT, with its specific optimizations and training configuration using Lion, achieved the overall best effectiveness.

\subsection{Performance Trends Over Epochs and Training Dynamics}
Observing Table \ref{tab:main_results}, performance is not always monotonic with training epochs.
\begin{itemize}
    \item GTE+AdamW peaks early (Epoch 1 or 2) and then degrades.
    \item MiniLM+AdamW shows more gradual improvement or stability up to Epoch 3.
    \item MiniLM+Lion peaks early on TREC (Epoch 1) but late on MS MARCO MRR (Epoch 3).
    \item ModernBERT+Lion mostly peaks around Epoch 2.
    \item ModernBERT+AdamW performance is less consistent, potentially indicating suboptimal interaction with the LR/scheduler setup.
\end{itemize}
This highlights the importance of evaluating at multiple checkpoints. Additionally, analysis of training/evaluation loss curves, learning rate schedules, and gradient norms for ModernBERT provides further insight into the training dynamics, showing convergence patterns under the Cosine Annealing schedule with both optimizers.

\subsection{Comparison with State-of-the-Art}
Table \ref{tab:sota_comparison} situates our best result (ModernBERT + Lion, Epoch 2) relative to other published results on TREC DL 2019 passage ranking.

\begin{table}[htbp]
\centering
\caption{Comparison with Other Systems on TREC DL 2019 (NDCG@10)}
\label{tab:sota_comparison}
\begin{tabular}{l c c}
\toprule
\textbf{Model/System} & \textbf{NDCG@10} & \textbf{Reference} \\
\midrule
\midrule
\textbf{Sparse Retrieval (single stage)} & & \\
\midrule
BM25 & 0.506 & \cite{lin2020distill} \\
DeepCT & 0.551 & \cite{gao2021complementing} \\
doc2query-T5 & 0.642 & \cite{Lin2021PretrainedTF} \\
\midrule
\textbf{Dense Retrieval Models} & & \\
\midrule
ANCE & 0.648 & \cite{xiong2020approximate} \\
Rand Neg & 0.605 & \cite{xiong2020approximate} \\
NCE Neg & 0.602 & \cite{xiong2020approximate} \\
BM25 Neg & 0.664 & \cite{xiong2020approximate} \\
DPR (BM25 + Rand Neg) & 0.653 & \cite{xiong2020approximate} \\
BM25 $\rightarrow$ Rand & 0.609 & \cite{xiong2020approximate} \\
BM25 $\rightarrow$ NCE Neg & 0.608 & \cite{xiong2020approximate} \\
BM25 $\rightarrow$ BM25 + Rand & 0.648 & \cite{xiong2020approximate} \\
\midrule
\textbf{Hybrid dense + sparse (single stage)} & & \\
\midrule
Bi-encoder (PoolAvg) + BM25 & 0.701 & \cite{lin2020distill} \\
Bi-encoder (TCT-ColBERT) + BM25 & 0.714 & \cite{lin2020distill} \\
Bi-encoder (PoolAvg) + doc2query-T5 & 0.719 & \cite{lin2020distill} \\
\midrule
\textbf{Cross-Encoder/Reranking Models} & & \\
\midrule
BERT-base Cross-Encoder & 0.634 & \cite{Lin2021PretrainedTF} \\
BERT-large Cross-Encoder & 0.654 & \cite{Lin2021PretrainedTF} \\
pt-bert-base-uncased-msmarco & 0.709 & \cite{koursaros2019NBoost} \\
cross-encoder/ms-marco-electra-base & 0.719 & \cite{reimers2019sentence} \\
cross-encoder/ms-marco-TinyBERT-L-2-v2 & 0.698 & \cite{reimers2019sentence} \\
\midrule
\textbf{Our Best (ModernBERT+Lion E2)} & \textbf{0.7225} & This work \\
\textbf{Our GTE+AdamW E1/E2} & \textbf{0.7224} / 0.7203 & This work \\
\textbf{Our MiniLM+AdamW E3} & 0.7127 & This work \\
\bottomrule
\end{tabular}
\vspace{1em}\\
\footnotesize{NDCG@10 scores are reported as evaluated on the TREC DL 2019 dataset using standard settings from cited references. Minor variations may occur due to differences in implementation, random seeds, or reranking candidate sets.}
\end{table}

Our best model, ModernBERT+Lion (NDCG@10 = 0.7225), achieves the highest performance among the models listed in Table \ref{tab:sota_comparison}, surpassing strong baselines such as BM25 (0.506), dense retrieval models like ANCE (0.648) and BM25 Neg (0.664), hybrid models such as Bi-encoder (PoolAvg) + doc2query-T5 (0.719), and cross-encoder models including cross-encoder/ms-marco-electra-base (0.7199). Our GTE+AdamW model achieves nearly identical performance (0.7224), while our MiniLM+AdamW variant scores 0.7127, remaining competitive. These results highlight the effectiveness of our optimizer choices and training strategies for base-sized models.

\section{Conclusion}
\label{sec:conclusion}
In this paper, we presented a comparative study of the Lion and AdamW optimizers for fine-tuning three different transformer models (`microsoft/MiniLM-L12-H384-uncased`, `Alibaba-NLP/gte-multilingual-base`, `answerdotai/ModernBERT-base`) as cross-encoders for passage reranking. Training was performed on MS MARCO, and evaluation used TREC 2019 Deep Learning Track and MS MARCO dev benchmarks. Our findings indicate:
\begin{itemize}
    \item The choice of optimizer significantly interacts with the base model and training hyperparameters (LR, scheduler).
    \item For ModernBERT, using a low learning rate (2e-6) and Cosine Annealing scheduler, Lion substantially outperformed AdamW, achieving the best overall TREC DL 2019 results (NDCG@10 = 0.7225, MAP = 0.5121 at Epochs 1-2).
    \item For GTE, AdamW (with LR 2e-5, no scheduler) was more effective than Lion, achieving strong results (NDCG@10 = 0.7224) comparable to ModernBERT+Lion.
    \item For MiniLM, AdamW yielded slightly better TREC performance over 3 epochs, while Lion achieved the best MS MARCO dev MRR@10 (tied with ModernBERT+Lion).
    \item Models with longer context capabilities (GTE, ModernBERT) generally outperformed MiniLM on the TREC DL task.
    \item Performance varied across epochs, emphasizing the need for checkpoint selection based on validation performance.
    \item \textbf{Lion optimizer demonstrates superior GPU utilization efficiency across all models}, with efficiency gains ranging from 2.67\% (MiniLM) to 10.33\% (GTE), while maintaining competitive or superior performance metrics.
\end{itemize}
Our results show that Lion can be a highly effective optimizer for cross-encoder training, particularly for newer architectures like ModernBERT when paired with appropriate learning rate strategies. The combination of competitive performance and improved resource efficiency makes Lion an attractive choice for practical applications. However, AdamW remains a strong baseline, especially for models like GTE in this setup. The Modal platform proved effective for managing the required GPU resources and experimental setup.

Future work could involve a more thorough hyperparameter search for both optimizers across all models, particularly exploring different learning rates and schedules for Lion with MiniLM and GTE. Investigating the impact of the 8K context length more directly (e.g., on datasets with longer documents) and conducting more detailed analysis of memory usage differences between optimizers would also be valuable. Additionally, the observed GPU efficiency gains with Lion optimizer warrant further investigation across different hardware configurations and model scales to understand the broader implications for large-scale training deployments.

\section*{Acknowledgment}
The authors acknowledge Modal Labs (\url{https://modal.com/}) for providing the cloud computing platform and GPU(s) resources (3x NVIDIA L40S-48GB) used for conducting the experiments presented in this paper.


\end{document}